\documentclass[fleqn]{iac}

\usepackage[table]{xcolor}
\usepackage[nolist, nohyperlinks]{acronym}
\usepackage[hyphens]{url}
\usepackage[hidelinks]{hyperref} 
\usepackage{multirow}
\hypersetup{breaklinks=true}
\usepackage{longtable,tabularx}
\usepackage{amsmath}
\usepackage{physics}

\usepackage{caption}
\usepackage{subcaption}
\usepackage{enumitem}
\usepackage{cite}
\usepackage{amsmath,amssymb,amsfonts}
\usepackage{algorithmic}
\usepackage{graphicx}
\usepackage{textcomp}
\usepackage{xcolor}
\usepackage{tikz}
\usepackage{subcaption}
\usepackage{booktabs}
\usepackage[noend]{algorithm2e}
\usepackage{siunitx}

\SetKwComment{Comment}{/* }{ */}
\RestyleAlgo{ruled}

\makeatletter
\def\tagform@#1{\maketag@@@{(\ignorespaces#1\unskip\@@italiccorr)}}
\makeatother

\usepackage{makecell}

\definecolor{lightbg}{rgb}{0.9, 0.9, 0.9}
\definecolor{lightgray}{rgb}{0.95, 0.95, 0.95}

\usepackage{hyperref}
\begin{document}

\IACpaperyear{23}
\IACpapernumber{B2.2.2}
% \IACconference{73}
% \IAClocation{Baku, Azerbaijan, 2-6 October 2023}
% \IACcopyrightA{2023}{the International Astronautical Federation (IAF)}

\title{
    Toward Autonomous Cooperation in Heterogeneous Nanosatellite Constellations Using Dynamic Graph Neural Networks
}

\IACauthor{{G. Casadesus Vila$^a$}, J.A. Ruiz-de-Azua$^b$, E. Alarcon$^c$}{$^a$Stanford University, $^b$i2CAT Foundation, $^c$Universitat Politecnica de Catalunya (UPC)}

\abstract{
    The upcoming landscape of Earth Observation missions will be defined by networked heterogeneous nanosatellite constellations required to meet strict mission requirements, such as revisit times and spatial resolution.
    However, scheduling satellite communications in these satellite networks through efficiently creating a global satellite Contact Plan (CP) is a complex task, with current solutions requiring ground-based coordination or being limited by onboard computational resources.
    The paper proposes a novel approach to overcome these challenges by modeling the constellations and CP as dynamic networks and employing graph-based techniques.
    The proposed method utilizes a state-of-the-art dynamic graph neural network to evaluate the performance of a given CP and update it using a heuristic algorithm based on simulated annealing.
    The trained neural network can predict the network delay with a mean absolute error of 3.6 minutes.
    Simulation results show that the proposed method can successfully design a contact plan for large satellite networks, improving the delay by 29.1\%, similar to a traditional approach, while performing the objective evaluations 20x faster.
}

\maketitle

% \printglossary[title=Acronyms/Abbreviations, type=\acronymtype]
% \begin{description}
% \item[]    
% \end{description}

\section{Introduction}

\subsection{Motivation}

% Easy access to space
The current New Space era has been characterized by increased access to space, driven significantly by the advent of small satellites and reduced launch costs, with 3,500 Earth Observation (EO) satellites to be launched over the next decade~\cite{Kodheli2021a,euroconsult2022earth}.
Space agencies, academic institutions, and private companies have adopted small satellite architectures for varied applications, ranging from scientific research and training to technology demonstrations and other space-based industrial applications.
Now, the space sector stands on the cusp of another leap, transitioning towards assembling and operating large, network-enabled heterogeneous nanosatellite constellations that will enable new Earth observation capabilities.

% The benefits of sharing
Incorporating inter-satellite links (ISLs) in EO constellations brings the immense advantages of inter-satellite collaboration, allowing them to meet stringent mission requirements such as real-time high-resolution global mapping~\cite{Curzi2020}.
% By facilitating reliable data transfers between satellites, ISLs ensure that Earth observation is continuous and global.
Seamless data-sharing capabilities across remote regions empower EO satellite constellations to create large observational networks.
Beyond the data, satellites can leverage unused onboard resources from their peers, increasing the scientific return.
This ecosystem of sharing and collaboration, enabled by the agile operation of a large nanosatellite constellation with inter-satellite communications, allows for the maximization of resources, enhanced operational efficiency, and a collaborative approach to achieving mission objectives.

% Delay/disruption tolerant networks
While collaboration and sharing offer numerous benefits, satellite constellations are subject to resource constraints and evolving network topologies, categorizing them as delay/disruption tolerant (DTN) networks.
In contrast to traditional end-to-end networks, DTNs are characterized by intermittent connectivity, long and variable delays, and high error rates.
In DTNs, routes between nodes need to be defined over long timespans, requiring new approaches to network management~\cite{Fall2003}.

% The need for autonomy
The complexity of managing large EO networks underscores the pressing need for autonomy~\cite{Araguz2018, Curzi2020}.
One of the critical problems in this context is how to efficiently schedule satellite communications given their communication capabilities, available onboard resources, downlink opportunities, and mission constraints.
Generating this schedule is known as the contact plan design (CPD) problem.
Traditional approaches, often centralized and reliant on human intervention, are becoming untenable, especially when dealing with constellations managed by diverse providers, as will be the case for the upcoming EO landscape.
In light of these challenges, moving towards autonomous systems is not just beneficial but imperative.
Autonomy can significantly enhance the constellation robustness, efficiency, and cost of operations, allowing satellites to make real-time decisions based on evolving mission needs.

\subsection{Literature Review}

% The Contact Plan Design problem
Due to the increasing interest in inter-satellite connectivity, the CPD problem has received significant attention over the last few years~\cite{Fraire2015}. Traditionally, the problem has been solved with three methodologies involving different levels of information and complexity: topology-based, traffic-based, and route-based.

% Topology aware - too simple, does not include information
% Traffic  aware - too complex, not enough information
Topology-based CPD methods are grounded on the deterministic nature of satellite orbital motion and only require the basic assumption of a known network topology. While they provide a simple and scalable solution, their performance is limited. On the other end of the spectrum, traffic-aware CPD methods consider the traffic demand between satellites and aim to optimize the network performance. However, these methods are computationally expensive and require a large amount of information, which is not always available. In this context,
the authors in~\cite{Wang2016} formulate the CPD problem as a flow optimization problem and solve it using stochastic optimization. By further modeling the satellite resources such as buffer and battery capacity, the authors~\cite{Zhou2017} propose a primal decomposition method and heuristic algorithm to solve the CPD problem while considering different mission requirements.

% Route aware - right information, not scalable
Route-aware CPD methods are a middle ground between the two previous approaches, requiring less information than traffic-aware methods while providing better performance than topology-aware methods.
Several works~\cite{Fraire2015c,Fraire2017a} proposed using Contact Graph Routing (CGR). This routing algorithm uses a contact graph to represent the contact plan and computes the best route between two nodes using a modified Dijkstra search, together with metaheuristic algorithms such as Simulated Annealing (SA) and Genetic Algorithms (GA) to maximize all-to-all route delay, provide fairness, and minimize undelivered traffic.
The authors in~\cite{Yan2015} also resort to SA and include link constraints to preserve diversity in navigation satellite constellations.
Under strong assumptions regarding the contact topology, recent work~\cite{Shi2018} focuses on multi-layer satellite networks.
These works, however, consider single missions and small homogeneous constellations, often in the context of ground-based centralized coordination.

% Research gap - distributed, scalable, heterogeneous methods for large constellations
While current studies cover different CPD methodologies, available information, and performance requirements, they fall short of addressing the complexity of large heterogeneous constellations.
In particular, they do not sufficiently study how the CP can be optimized onboard for large constellations managed by different entities, targeting computational performance and scalability.
Furthermore, they do not consider the dynamic nature of the network, limiting their scope to homogeneous constellations of reduced size.

% Graph neural networks for communications - not changing topology
We focus on learning-based methods to address this gap and target a good balance between performance and scalability.
Satellite and, more generally, communication networks comprise many fundamental components naturally represented in graph structures, such as the network topology or routing.
In fields where data is represented as dynamic graphs, dynamic graph neural networks (DGNN) have shown outstanding results~\cite{Skarding2020}.
The success of such applications, including delay and traffic prediction in wireless networks~\cite{Suarez-Varela2022a}, motivates their use for the CPD problem.

\subsection{Paper Objective}

This paper presents a method to model communication networks with dynamics topologies using dynamic graphs and employs DGNNs for network-level metrics computation.
We present a DGNN architecture for dynamic network modeling that integrates the spatial processing of Graph Neural Networks (GNN) with Recurrent Neural Networks (RNN) temporal processing.
This DGNN is trained to assess latency in satellite constellations represented as dynamic graphs and optimize contact plans using a simulated annealing-based metaheuristic algorithm.
Though our primary focus is on the CPD problem, we see this approach as a foundation for utilizing dynamic graph-based learning in managing heterogeneous satellite constellations.

\subsection{Overview}
The remainder of the paper is organized as follows: Section~\ref{sec:formulation} formulates the CPD problem with the implications introduced by heterogeneous EO constellations, Section~\ref{sec:methodology} describes the proposed CPD method with the use of DGNN for the objective function learning, Section~\ref{sec:results} shows the results of applying the method to different use cases. Finally, Section~\ref{sec:conclusion} remarks the conclusions of this work and possible future research directions.

% ********************************************************************
% ********************************************************************
\section{Problem Statement}\label{sec:formulation}
% ********************************************************************
% ********************************************************************

In this section, we formulate the CPD problem as a constrained optimization problem, similar to previous works~\cite{Yan2015, Fraire2015c}.
Two key differences are that we do not reduce the constellation to a finite set of periodic states due to the scale and heterogeneity of our target constellations and that we include source and destination traffic information for different types of nodes, namely satellites and ground stations.

\subsection{System model}

% Satellites, ground stations, and comms needs
We consider a non-gostationary orbit constellation with a set of satellites, $\mathcal{S}=\{1, 2, \dots, N_s\}$, that can communicate with each other and a set of ground stations, $\mathcal{G}=\{1, 2, \dots, N_g\}$, at discrete time steps $t \in \mathcal{T} = \{1, 2, \dots, N_t\}$. We assume that each satellite $i$ needs to communicate with a subset of satellites and grounds station $\mathcal{C}_i \subset \mathcal{S} \cup \mathcal{G}$, which is known a priori.

% Topology
We model the communication opportunities between satellites and ground stations, which depend on their communication capabilities and operational constraints, as a sequence of visibility matrices $V = (V_1, \dots, V_{N_t})$, where $V_t \in \mathbb{R}^{(N_s+N_g) \times (N_s+N_g)}$, and $V_t(i, j)=1$ if communication between node $i$ and $j$ is feasible at time $t$ and $V_t(i, j)=0$ otherwise.
Therefore, a visibility matrix represents the communication feasibility between satellites and ground stations at different time instances, capturing their communication constraints, e.g., the inter-satellite range for satellites and minimum elevation angle for ground stations, all assumed to be known a priori.

% Contact plan
We define the contact plan to be generated as a sequence of matrices $U = (U_1, \dots, U_{N_t})$, where $U_t \in \mathbb{R}^{(N_s+N_g) \times (N_s+N_g)}$ is the contact plan matrix at time $t$.
The contact plan matrix $U_t$ represents the scheduled communication links between nodes at time $t$, where $U_t(i, j)=1$ if there is a link to be established between nodes $i$ and $j$ at time $t$ and $U_t(i, j)=0$ otherwise.

% Constraints: bidirectional links, max number of links
We assume that satellites and ground stations establish bidirectional links, i.e., $U_t(i, j)=U_t(j, i)$, and that the maximum number of links that a satellite can establish with other satellites and ground stations is $M_s$ and $M_g$, respectively.
These constraints are motivated by mission requirements and limit the number of links a satellite can establish during the period for which the contact plan needs to be generated.

% Objective
For a given contact plan, the best delivery time (BDT) between two nodes $i$ and $j$ at time $t$, $d(i, j, t)$, is the time that it takes for a message sent a time $t$ from satellite $i$ to reach satellite $j$, accounting only for the changing topology of the network.
That is, without taking into account other traffic or link capacity.
The objective of the CPD problem is to find the contact plan that minimizes the BDT between pairs of nodes.

\begin{align}
    \mbox{minimize} \quad
     & F = \sum_{t\in\mathcal{T}} \sum_{i\in\mathcal{S}} \sum_{j\in\mathcal{C}_i} d(i, j, t) \label{obj} \\
    \mbox{subject to} \quad
     & U_t(i, j) = U_t(j, i),
     & \forall i, j, t \label{eq:constr_sym}                                                             \\
     & U_t(i, j) \leq A_t(i, j),
     & \forall i, j, t \label{eq:constr_vis}                                                             \\
     & \sum_{t\in\mathcal{T}} \sum_{j\in\mathcal{S}} U_t(i, j) \leq M_s,
     & \forall i \label{eq:constr_sats}                                                                  \\
     & \sum_{t\in\mathcal{T}} \sum_{j\in\mathcal{G}} U_t(i, j) \leq M_g,
     & \forall i \label{eq:constr_gs}
\end{align}

where $U_t(i, j)$ is the optimization variable encoding the contact plan. Constraint~\eqref{eq:constr_sym} enforces bidirectional links,~\eqref{eq:constr_vis} ensures that the contact plan is feasible under the satellite topology, and~\eqref{eq:constr_sats} and~\eqref{eq:constr_gs} limit the number of links that a satellite can establish with other satellites and ground stations, respectively.

% ********************************************************************
% ********************************************************************
\section{Methodology}\label{sec:methodology}
% ********************************************************************
% ********************************************************************

In this section, we present the proposed method for the CPD problem, which consists of two main components: a heuristic algorithm based on simulated annealing for contact plan design and a dynamic graph neural network for evaluating the objective function.

% ********************************************************************
\subsection{Simulated Annealing}\label{sec:cp_desgin}
% ********************************************************************

% The need for a metaheuristic approach
Considering that the contact plan problem defined in the previous section would rapidly become computationally intractable, we use a heuristic algorithm based on simulated annealing (SA), a stochastic optimization method that mimics the slow cooling process in metallurgy to gradually reduce the search space and decrease the likelihood of accepting suboptimal solutions~\cite{bertsimas1993simulated}.
This approach has been demonstrated to be effective in topology design problems~\cite{Yan2015, Fraire2015}, enabling the acquisition of sub-optimal solutions. Algorithm~\ref{alg:CPDSA} (Appendix~\ref{app:algorithms}) presents the contact plan design algorithm based on SA.

We begin by creating an initial contact plan, from which we obtain a new contact plan while ensuring constraint satisfaction by adding or removing links.
Then, we calculate and compare the objective functions of the new and the current contact plan. If the objective function of the new contact plan is improved, we accept the new contact plan and reduce the temperature $T$.
Otherwise, we accept the new contact plan with a probability $P=\exp \left(-(F_{new}-F_{curr})/T\right)$, where $F_{new}$ and $F_{curr}$ are the objective functions of the new and the current contact plans, respectively.
If the new contact plan is accepted, we reduce the temperature $T$ by a cooling rate $r$. The algorithm stops when a set number of iterations $N_{it}$ is reached.

The best delivery time between two nodes $d(i, j, t)$ is typically calculated using algorithms such as Contact Graph Dijkstra Search (Algorithm~\ref{alg:CGDS}, Appendix~\ref{app:algorithms}), a modified Dijkstra search that finds the best route through a contact graph~\cite{Fraire2021}, which is a graph representation of the contact plan that allows computing routes between nodes. However, since the above method can be computationally expensive for large networks, the following sections present a learning-based approach based on DGNNs.

%The main advantage of expressing the contact plan as a contact graph data structure is that it can be used as input for traditional shortest-path algorithms. % such as the one presented in this work.

% ********************************************************************
\subsection{Dynamic Graph Neural Networks}
% ********************************************************************

\begin{figure*}[tbp]
    \centering
    \includegraphics[width=0.9\linewidth]{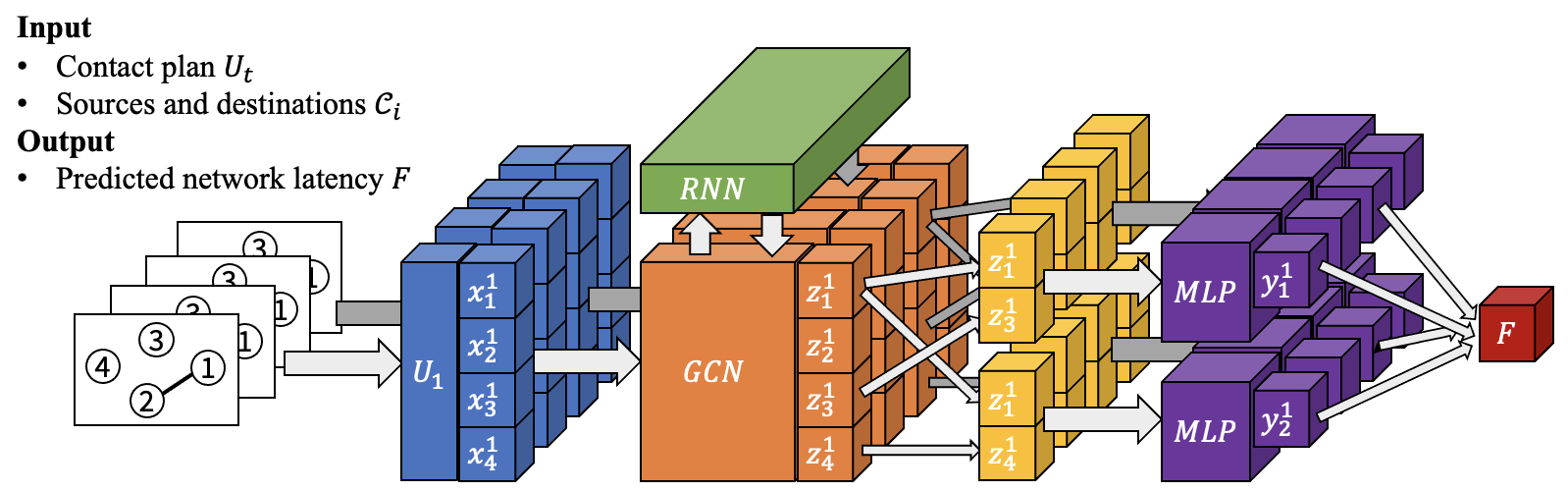}
    \caption{Dynamic graph neural network architecture for predicting latency in satellite constellation networks.}\label{fig:dgnn}
\end{figure*}

% Motivate dynamic graph neural networks
The objective function is evaluated by inputting a contact plan encoded by a sequence of matrices $U = (U_1, \dots, U_{N_t})$.
This sequence can be interpreted as a dynamic graph, where each matrix $U_t$ represents the adjacency matrix at a given time step. Therefore, we propose using a dynamic graph neural network (DGNN) to learn the objective function and use it to evaluate the contact plan performance.
More specifically, we represent the contact plan $U$ as a sequence of graphs $G=(G_1, \dots, G_{N_t})$, with $G_t=(\mathcal{V}, \mathcal{E}_t)$, where $\mathcal{V}=\mathcal{S} \cup \mathcal{G}$ is the set of nodes and $\mathcal{E}_t$ is the set of edges at time $t$. Each edge $e_{ij} \in \mathcal{E}_t$ represents a link between nodes $i$ and $j$ at time $t$, as specified by the contact plan matrix at that time instance $U_t(i, j)$.

% DGNNs
The selected DGNN model to learn the objective function $F$ is EvolveGCN~\cite{Pareja2019}. One of its main features is handling the addition and removal of nodes after training, overcoming the limitation of other methods in learning these irregular behaviors.
The authors propose using a Recurrent Neural Network (RNN) to regulate a Graph Convolutional Network (GCN) model (i.e., network parameters) at different time steps.
This approach effectively performs what is known as model adaptation by focusing on the model itself rather than the node embeddings.
Therefore, the change of nodes poses no restriction, making the model sensitive to new nodes without historical information.

As in the original paper's nomenclature~\cite{Pareja2019}, we will use subscript $t$ to denote the time index, superscript $l$ to denote the GCN layer index, $n$ for the number of nodes---in our case $n=N_s+N_g$. At a time step $t$, the input data to the model consists of the pair $(A_t \in \mathbb{R}^{n\times n}, X_t\in \mathbb{R}^{n\times d})$, where the first element is the graph adjacency matrix---in our case, obtained from the contact plan $U_t$---and the second is the matrix of input node features.
Specifically, each row of $X_t$ is a vector of node $d$ features.

The GCN consists of $L$ layers of graph convolution, which includes a neighborhood aggregation that combines information from neighboring nodes. At time $t$, the $l$-th layer takes as input the adjacency matrix $A_t$ and the learnable node embedding matrix $H^{(l)}_t$, and uses the learnable weight matrix $W^{(l)}_t$ to update the node embedding matrix to $H^{(l+1)}_t$
\begin{align}
    H^{(l+1)}_t & = \text{GCONV}\pqty{A_t, H^{(l)}_t, W^{(l)}_t} \\
                & := \sigma\pqty{\hat{A}_t H^{(l)}_t W^{(l)}_t}
\end{align}
where $\hat{A}_t$ is the normalized adjacency matrix and $\sigma$ is the activation function. The initial embedding matrix is obtained from the node features, $H^{(0)}_t = X_t$.

The central component of the model is the update of the weight function $W^{(l)}_t$ at each time step. The weights are updated by an rnnRNN that takes as input the node embedding matrix $H^{(l)}_t$ and the previous weight matrix $W^{(l)}_{t-1}$ and outputs the updated weight matrix $W^{(l)}_t$ as
\begin{equation}
    W^{(l)}_t = \text{RNN}\pqty{H^{(l)}_t, W^{(l)}_{t-1}}
\end{equation}

Combining the graph convolution with the recurrent architecture, the authors define the evolving graph convolution unit, which is the basic building block of the EvolveGCN model.

Since we assume the satellites and ground stations that each satellite communicates with, $\mathcal{C}_i$, as known, for each time step, we create link embeddings by concatenating the embeddings of the source and destination nodes.
That is, we concatenate the embeddings of the final layer at different time steps, $H^{(L)}_1, \dots, H^{(L)}_{N_t}$, to compute the objective function $F$ using a fully connected neural network with a single output and average pooling.

% ********************************************************************
% ********************************************************************
\section{Results}\label{sec:results}
% ********************************************************************
% ********************************************************************
In this section, we first present the experimental setup, including the parameters of the DGNN model and training. Then, we present the results of the contact plan design based on simulated annealing using a DGNN for evaluating the objective.

\subsection{DGNN Model}
Due to the time required to compute the objective function for large constellations and generate training data, we trained the final model for 16 hours using synthetic data corresponding to 30 satellites and 20 ground stations.
We performed hyperparameter tuning, in which we considered different activation functions and loss functions, namely $L_1$ and $L_2$, as well as the number of layers and size for the graph neural network, the recurrent neural network, and the fully connected neural network. We also considered different architectures, such as a standalone GCN, and different ways of processing the node embeddings to predict the objective function.

% feature
% "degree"
% loss
% "l1"
% lr
% 0.001
% model
% "EvolveGCN-O"
% n_hidden
% 64
% n_hidden_classifier
% 64
% n_layers
% 2

\begin{figure}[tbp]
    \centering
    \includegraphics[width=1.0\linewidth]{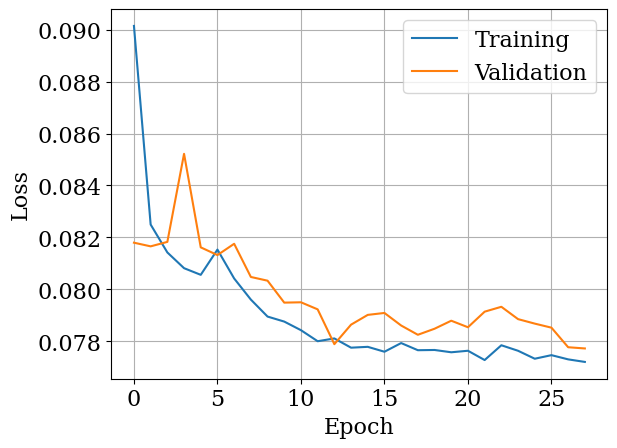}
    \caption{Training and evaluation loss. The model is trained for 16 hours using synthetic data corresponding to 30 satellites and 20 ground stations. Hyperparameters are selected using a grid search, including different activation and loss functions, as well as the number of layers and sizes.
    }\label{fig:loss}
\end{figure}

The best-performing model is EvolveGCN-O, consisting of 2 layers of graph convolution with 64 hidden units, a fully connected neural network with 2 layers of 64 hidden units, and a single output. The model is trained using the Adam optimizer with a learning rate of 0.001 and the $L_1$ loss function. The graph convolution layer has Randomized Leaky Rectified Linear Units (RReLu) as the activation function, and the fully connected neural network has Rectified Linear Units (ReLu) as the activation function. Figure~\ref{fig:loss} shows the training and validation loss for the selected model. The implementation uses the Deep Graph Library~\cite{wang2019deep} and PyTorch~\cite{paszke2019pytorch}.

\begin{figure}[tbp]
    \centering
    \includegraphics[width=1.0\linewidth]{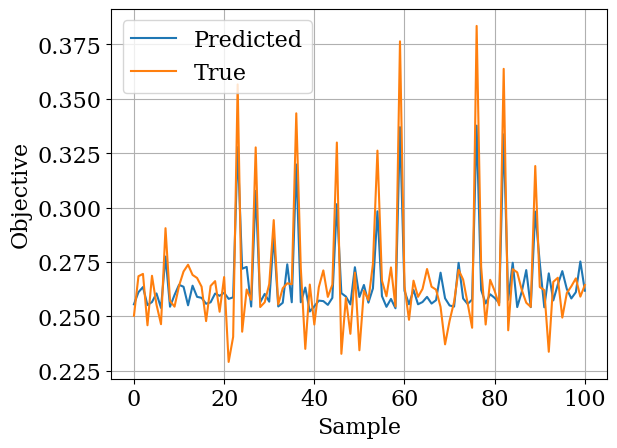}
    \caption{Predicted and true normalized BDT for 100 different contact plans. The model successfully identifies contact plans with worse objective values and achieves lower accuracies on contact plans with lower objectives. It predicts the BDT of a contact plan with a mean absolute error of 3.6 minutes.
    }\label{fig:pred_true}
\end{figure}

Figure~\ref{fig:pred_true} shows the predicted and true normalized BDT for 100 different contact plans. The model successfully identifies contact plans with worse objective values and achieves lower accuracies on contact plans with lower objectives.
This indicates that the model can learn the objective function and can be used to evaluate the performance of a given contact plan, which is the main objective of this work.
Moreover, we expect better performance at the beginning of the optimization, when the contact plan has higher objective values, and worse performance at the end when the contact plan has lower objective values.

\begin{figure}[tbp]
    \centering
    \includegraphics[width=0.95\linewidth]{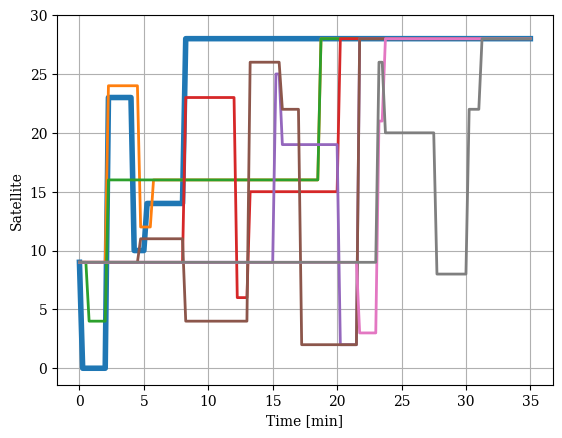}
    \caption{
        Possible routes from satellite 9 to satellite 28. The best route is colored in blue, passing through satellites 9, 0, 23, 10, 14, and 28, with a Best Delivery Time (BDT) of 8 minutes. The proposed DGNN can predict the average BDT of the network with a mean absolute error of 3.6 minutes.
    }\label{fig:reachable_satellites}
\end{figure}

Fig.~\ref{fig:reachable_satellites} shows the different routes from satellite 9 to satellite 28. The best route is colored in blue, passing through satellites 9, 0, 23, 10, 14, and 28, with a Best Delivery Time (BDT) of 8 minutes. The proposed DGNN can predict the average BDT of the network with a mean absolute error of 3.6 minutes. The different paths involving multiple satellites highlight the complexity of the objective function, which is not only dependent on the distance between the source and destination nodes but also on the dynamic topology of the network.

\subsection{Contact Plan Design Results}
We assume that each satellite is only equipped with a single ISL and that inter-satellite communication or downlink cannot occur simultaneously.
Regarding the optimization algorithm based on SA, we generate the initial random contact plan by solving a maximum matching problem, a well-known combinatorial optimization problem consisting of finding a maximum cardinality matching in a graph.
This readily enforces the symmetry and visibility constraints.
The constraints related to the maximum number of contacts per satellite are enforced by recursively deactivating links.

We implemented the contact plan design based on simulated annealing in Algorithm~\ref{alg:CPDSA} (Appendix~\ref{app:algorithms}) in Python.
To generate a new plan, we randomly activate and deactivate new links while always ensuring constraint satisfaction. The initial temperature is set to $T=10$, the cooling rate is set to $r=0.95$, and the number of iterations is set to $N_{it}=100$. It is essential to mention that when computing the BDT, we consider the delay from the average distance between satellite pairs.

\begin{figure}[t]
    \centering
    \includegraphics[width=1.0\linewidth]{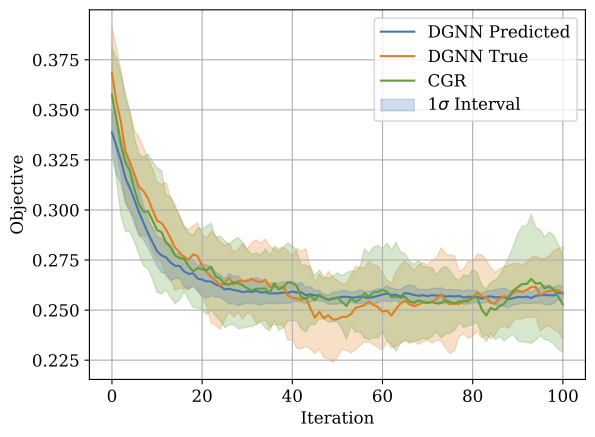}
    \caption{
        Objective improvement using the DGNN to evaluate the objective function. The predicted objective values using the trained model are shown in blue, and the values computed a posteriori using the CGR algorithm are shown in orange. The shaded area represents the standard deviation over 10 runs.
    }\label{fig:obj_vs_iter}
\end{figure}

% Objective Function
% CGR
% DGNN
% Initial
% 0.36±0.02
% 0.37±0.02
% Final
% 0.25±0.02
% 0.26±0.02
% Improvement
% 29.6%±7.3%
% 29.1%±7.34%
% Computing time (100 it)
% 19.1±0.8 min
% 0.9±0.2 min

\begin{table}[tbp]
    \footnotesize
    \centering
    \caption{
        Normalized average best delivery time (BDT) of the initial and optimized plans and computing times when using CGR and the proposed DGNN to compute the objective function. The results show the mean and standard deviation over 10 runs.
    }\label{tab:bdt}
    \begin{tabular}{lcc}
        \toprule
        \textbf{Objective Function} & \textbf{CGR}     & \textbf{DGNN}   \\
        \midrule
        Initial normalizd BDT       & 0.36$\pm$0.02    & 0.37$\pm$0.02   \\
        Final normalized BDT        & 0.25$\pm$0.02    & 0.26$\pm$0.02   \\
        Improvement                 & 29.6$\pm$7.3\%   & 29.1$\pm$7.34\% \\
        \midrule
        Computing time (100 it.)    & 19.1$\pm$0.8 min & 0.9$\pm$0.2 min \\
        \bottomrule
    \end{tabular}
\end{table}

Fig.~\ref{fig:obj_vs_iter} shows the decrease of the function value of the contact plan throughout the optimization using the trained model to predict the objective function.
We include both the values predicted by the model during optimization as well as the true values computed a posteriori.\@
The algorithm tends to converge after approximately 100 iterations, as better plans are not found and the temperature is reduced.
We also observe how the predicted values are close to the true values during the early phases of the optimization when plans have more significant delays.
However, the model's performance decreases as the optimization progresses, and the plans have lower delays, indicating that the model cannot generalize well to the entire range of delays.
This is shown by the large difference in standard deviation between the blue and the orange lines, representing the true values computed a posteriori.
The model tends to predict a mean value close to the true value but does not capture the variance of the objective function.

Nonetheless, the proposed method is able to design a contact plan for the simulation use case successfully.
Thanks to the initialization method based on constraint satisfaction and the implemented algorithm based on SA, we ensure that the contact plan is always feasible, allowing us to use the DGNN to evaluate the objective function.

Table~\ref{tab:bdt} shows the normalized average BDT, i.e., the objective function value of the initial and optimized plans, and the computing times when using CGR and the proposed DGNN to compute the objective function. We observe that the optimized contact plan using the DGNN is able to improve the BDT by 29.1\% compared to the initial contact plan, achieving similar results to optimizing using the CGR.\@
However, using the DGNN, we can perform the objective evaluations 20x faster than when using CGR, which significantly improves computational performance.
Although, as shown in Fig.~\ref{fig:obj_vs_iter}, the DGNN is not able to generalize well to the entire range of delays, it can provide a good approximation of the objective function and allow the optimization algorithm to find a good solution.

% ********************************************************************
% ********************************************************************
\section{Conclusions}\label{sec:conclusion}
% ********************************************************************
% ********************************************************************
Designing a contact plan for large heterogeneous satellite networks necessitates consideration of the trade-off between performance and scalability. In this paper, we initially present a problem formulation for the contact plan design, taking into account traffic source and destination nodes without making any assumptions about the periodicity of the evolving satellite network topology.

To reduce the computational burden of computing the objective function for large satellite networks and tackle the need to update the contact plan periodically, we propose modeling the objective function, i.e., the average best delivery time of the network, using a dynamic graph neural network. This allows resorting to a metaheuristic algorithm based on simulated annealing.

The presented architecture captures the spatial and temporal information of the satellite network using a hybrid model that involves graph convolution and a recurrent neural network, being able to predict the average BDT of the network with a mean absolute error of 3.6 minutes.
Simulation results show that the proposed method is able to design a contact plan for a large satellite network successfully, improving the BDT by 29.1\%, achieving similar results to optimizing using CGR but performing the objective evaluations 20x faster.

% Analyzing the proposed solution in scenarios with contact plan distribution
% Extend the model to predict network metrics with traffic information
% Extract network and topology information from the learned model

Future work will focus on analyzing the proposed solution in scenarios with contact plan distribution, extending the model to predict network metrics with traffic information of the satellites, and extracting network and topology information from the learned model.
While this work focuses on a specific instance of the contact plan design problem, we envision that the proposed learning-based method involving DGNNs opens the door to new solutions for autonomy in large, heterogeneous satellite networks.

\section*{Acknowledgment}
This work was supported by the ``laCaixa'' Foundation fellowship (ID 100010434, code LCF/BQ/EU21/11890112). This work was co-funded by the Government of Catalonia in the scope of the NewSpace Strategy for Catalonia and by the Spanish Ministry of Economic Affairs and Digital Transformation and the European Union--NextGeneration EU, in the framework of the Recovery Plan, Transformation and Resilience (PRTR) (Call UNICO I+D 5G 2021, ref. number TSI-063000-2021-5-6GSatNet-SS).

\appendix

\section*{A\@: Algorithms}\label{app:algorithms}
This appendix presents the algorithms used in this work, namely the contact plan design based on simulated annealing (Algorithm~\ref{alg:CPDSA}) and the contact graph Dijkstra search (Algorithm~\ref{alg:CGDS}).
\begin{algorithm}[h!]
    \caption{Contact plan design based on simulated annealing.~\cite{bertsimas1993simulated}}\label{alg:CPDSA}
    \KwData{adjacency matrices $A_t$, temperature $T$, cooling rate $r$, number of iterations $N_{it}$}
    \KwResult{contact plan $U$}
    $U \gets \mathrm{InitialContactPlan}(A)$ \;
    $F_{best} \gets \mathrm{ObjectiveFunction}(U)$ \;
    \For{$i \gets 1$ \KwTo~$N_{it}$}{
        $U' \gets \mathrm{GetNeighbor}(U)$ \;
        $F' \gets \mathrm{ObjectiveFunction}(U')$ \;

        \uIf{$F' \leq F_{best}$}{
            $U_{best} \gets U \gets U'$ \;
            $F_{best} \gets F'$ \;
        }
        \uElseIf{$\exp\left((F_{curr} - F_{new})/T\right) \leq \mathrm{rand}(0, 1)$}{
            $U \gets U'$ \;
            $T \gets r \cdot T$ \;
        }
        \Else{
            $U \gets U_{best}$ \;
        }
    }
\end{algorithm}\vspace{-1.76em}
\begin{algorithm}[h!]
    \caption{Contact Graph Dijkstra Search.~\cite{Fraire2021}}\label{alg:CGDS}
    \KwData{Root contact $C_{root}$, source $S$, destination $D$, contact plan $U$}
    \KwResult{Route $R_S^D$, and best delivery time $BDT$}
    $R_S^D \gets \{\}$ \;
    $C_{fin} \gets \{\}$ \tcp*{final contact}
    $BDT = \infty$ \tcp*{final arrival time}
    $C_{curr} = C_{root}$ \tcp*{current contact is root}

    \tcc{contact plan exploration loop}
    \While{True}{
        \tcc{contact review procedure}
        $C_{fin}, BDT = \mathrm{CRP}(U, C_{curr}, C_{fin}, BDT)$\;
        \tcc{contact selection procedure}
        $C_{next} = \mathrm{CSP}(U, C_{curr}, BDT)$\;
        \If{
            $C_{next}$}{$C_{curr} \gets C_{next}$
        }
        \Else{
            \textbf{break}}
    }
    \If{$C_{fin}\neq \{\}$}{
        $C = C_{fin}$\;
        \While{$C\neq C_{S,S}^{0,\infty}$}{
            $R_S^D.hops \gets \{C\}$\;
            $C = C.pred$ \tcp*{previous contact}
        }
        Compute $(R_S^D.tx\_win, R_S^D.volume)$\;
    }
\end{algorithm}

\bibliographystyle{ieeetr}
\bibliography{main}

\end{document}